\documentclass[10pt]{article}

\usepackage[utf8]{inputenc}
\usepackage[T1]{fontenc}
\usepackage[top=2.5cm, bottom=2.5cm, left=2.5cm, right=2.5cm]{geometry}
\usepackage{multicol}
\usepackage{amsmath}
\usepackage{seqsplit}
\usepackage{graphicx}
\usepackage{wrapfig}
\usepackage{caption}
\usepackage{color}
\usepackage[hidelinks]{hyperref}
 
\title{\Large{\textbf{X-ray holography of skyrmionic cocoons in aperiodic magnetic multilayers}}}

\author{M. Grelier$^1$, R. Battistelli$^2$, H. Popescu$^3$, F. Godel$^1$, A. Vecchiola$^1$, S. Collin$^1$, C. Léveillé$^3$, \\K. Bouzehouane$^1$, F. Büttner${^{2,4}}$, V. Cros$^1$, N. Jaouen$^3$, N. Reyren$^{1*}$}
\date{\textit{\small{$^1$ Unit\'e Mixte de Physique, CNRS, Thales, Universit\'e Paris-Saclay, 91767, Palaiseau, France.\\
$^2$ Helmholtz-Zentrum Berlin, 14109 Berlin, Germany.\\
$^3$Synchrotron SOLEIL, L’Orme des Merisiers, 91192, Gif-sur-Yvette, France.\\
$^4$ Augsburg University, 86159, Augsburg, Germany.\\
$^*$ email: nicolas.reyren@cnrs-thales.fr}}}

\begin{document}

\maketitle

\section*{ABSTRACT}

The development and characterization of three-dimensional (3D) topological magnetic textures has become an important topic in modern magnetism both for fundamental and technological perspectives. Among the novel 3D spin textures, skyrmionic cocoons have been successfully stabilized in magnetic multilayers having a variable thickness of the ferromagnet in the vertical direction of the stack. These ellipsoidal 3D magnetic textures remain vertically confined in a fraction of the total thickness while coexisting with fully columnar skyrmions. Here, we use X-ray holography with about 15\,nm lateral resolution to investigate how their properties depend on the field and temperature. We observe circular objects with different amplitude of contrast which evidences the presence of different 3D objects located in various vertical parts of the multilayer. Moreover, we witness during out-of-plane cycling an attractive interaction between cocoons located at various heights, mainly due to the stray field, which impacts their horizontal positioning. The X-ray holography measurements also allow to determine the size of the cocoons at remanence which, at room temperature, possess diameter close to 100\,nm in average. Combining this transmission technique with magnetic force microscopy and micromagnetic simulations gives a precise insight into the 3D distribution of the magnetization which demonstrate the 3D nature of skyrmionic cocoons.

\section*{INTRODUCTION}

Magnetic topological textures have been identified as promising information bits for the development of next-generation spintronics devices, especially in terms of stability and energy consumption. A pioneer example of such objects remains the magnetic skyrmion \cite{roessler2006spontaneous,heinze2011spontaneous,nagaosa2013topological,fert2017magnetic}, a two-dimensional (2D) whirl of the magnetization that is in most cases stabilized by the Dzyaloshinskii-Moriya interaction (DMI), arising from broken inversion symmetry and large spin-orbit coupling. Skyrmions can be observed in chiral magnets or in magnetic multilayers where they acquire a vertical tubular shape, however in some cases with an hybrid chirality over the z-direction \cite{buttner2018theory,legrand2018hybrid}. Despite the potential of skyrmions and other 2D topological textures, a new interest has arisen for more complex quasi-particles that display variations over the thickness, i.e., three-dimensional (3D) objects \cite{gobel2021beyond}. The added dimensionality can lead to more intricate magnetic distribution which allows the range of perspective applications to be broadened \cite{fernandez2017three, raftrey2022road}. As a consequence, an ever-growing set of 3D solitons is already developing. For instance, in non-centrosymmetric crystals, magnetic bobbers that correspond to half tubular skyrmions ending with a Bloch point burried inside the ferromagnetic film have been recently observed  \cite{redies2019distinct,zheng2018experimental}, as well as the topologically trivial dipole strings \cite{muller2020coupled,leonov2018homogeneous} in bulk compounds. In magnetic multi-layered thin films, truncated skyrmions \cite{mandru2020coexistence,yildirim2022tuning}, or even hopfions \cite{kent2021creation,liu2018binding} have been reported. A recent newcomer is the skyrmionic cocoon \cite{grelier2022three}, so named due to its peculiar ellipsoidal shape. We recently stabilized such cocoons in aperiodic multilayers with a complex distribution of the magnetic interactions, making use of the tunability of such structures. Skyrmionic cocoons only reside in a fraction of the multilayers and this vertical confinement can be tuned either by optimizing the multilayer architecture or by using an external magnetic field. Interestingly, they can coexist with more usual textures, such as columnar skyrmions, an attractive feature both for the explanation of fundamental phase transitions between different topological states \cite{buttner2021observation} and for potential applications such as racetrack memories based on different types of 3D textures \cite{zheng2017experimental}. 

In this study, we report a non-perturbative observation of skyrmionic cocoons as well as their field dependent behaviour measured via X-ray holography with extended reference by autocorrelation linear differential operator (HERALDO) \cite{guizar2008direct} along with X-ray Fourier transform holography (FTH) experiments \cite{eisebitt2004lensless}. HERALDO and FTH are lensless high-resolution transmission techniques that record the hologram resulting from the interferences between an object hole and a reference aperture, a slit for the former and a hole for the latter. Those create a reference wave that encodes the phase so that a simple Fourier transform directly allows to recover a real-space image of the actual magnetic configuration of the sample. The spatial resolution is typically limited by the size of the reference aperture as well as by the numerical aperture of the setup. To improve the contrast and the resolution, one can additionally make use of phase retrieval algorithms \cite{marchesini2007invited} for the image reconstruction. The experiments have been performed on the COMET \cite{popescu2019comet} end-station of the SEXTANTS beamline located at the SOLEIL synchrotron using a transmission setup to measure first the circularly left (CL) polarization followed by the circularly right (CR) one at the Co L$_3$ edge (778.8 eV). The vertical component of the magnetization averaged over the material thickness can then be extracted from the difference between the real-space reconstructions of the two polarization images, which informs about the thickness extent of the 3D magnetization textures. As holography is a transmission technique, it allows to probe the full magnetization distribution, contrarily to more accessible techniques like magnetic force microscopy (MFM) which is more surface sensitive. The comprehensive field evolution study of skyrmionic cocoons, including the holography measurements along with MFM and micromagnetic simulations, allows to better characterize these new 3D textures.	

\section*{Results}

\subsection*{Multilayers hosting skyrmionic cocoons}

We recently show that skyrmionic cocoons can be stabilized in magnetic multilayers displaying variable thickness of the ferromagnetic films thanks to the induced inhomogeneous distribution of the magnetic interactions \cite{grelier2022three}. The architecture considered in the HERALDO study is detailed in Fig. 1a, notably showing the thickness evolution of the successive trilayers Pt/Co/Al (layer number is shown in horizontal axis). This schematic drawing evidences three distinguishable parts: the top and bottom ones in which the Co thickness varies from layer to layer, that we later on call the gradient parts, and the middle ones (between layer number 38 and 84) where the Co thickness is fixed at 1\,nm to have a strong perpendicular magnetic anisotropy (PMA). The sample imaged with FTH was slightly modified as the thickness of the Pt layers in the top gradient have been altered (see Methods for the precise structures). The multilayers have been grown by sputtering deposition on the top side of a Si$\rm{_3}$N$\rm{_4}$ membrane with a back side covered with a 1.1 µm-thick gold mask layer that was milled away with focused ion beam, yielding for HERALDO (resp. FTH) a 2\,µm diameter (resp. 800\,nm) round object hole. In Fig. 1b, the alternating gradient force magnetometry (AGFM) curve displays the magnetization $M$ normalized by the saturation magnetization $M_s$, for an out-of-plane (OOP) magnetic field sweeping from positive to negative. In conjugation with micromagnetic simulations performed with Mumax3 \cite{vansteenkiste2014design}, it is possible to identify the various magnetic phases hosted by the multilayer based on the transitions in the signal derivative, following our method described in \cite{grelier2022three}. Starting at high positive field, the magnetization points uniformly along the field direction. Upon the decrease of the field, around 275\,mT, skyrmionic cocoons begin to nucleate in the gradient parts of the structure while the middle layers are still pointing up. At 125\,mT, a jump is measured in the magnetization, which corresponds to the formation of domains that extend over the whole thickness of the multilayer, which we will call 3D stripes. At remanence, a coexistence is observable between those 3D stripes and skyrmionic cocoons that are only present in the top and bottom outer parts with the Co gradients (see Fig. 1c). Decreasing more the magnetic field can transform the 3D stripes into columnar skyrmions and ultimately provokes the disappearance of textures in the strong PMA layers, leaving again only cocoons behind (between -260 and -325\,mT) that will eventually be annihilated near -325\,mT. In Fig. 1c, a vertical cut of the vertical component of the magnetization ($m_z$) obtained with micromagnetic simulations illustrates the spatial vertical extent of each of the textures at remanence which can evolves significantly with the magnetic field.

\begin{figure}[htb!]
\includegraphics[width = 1\textwidth]{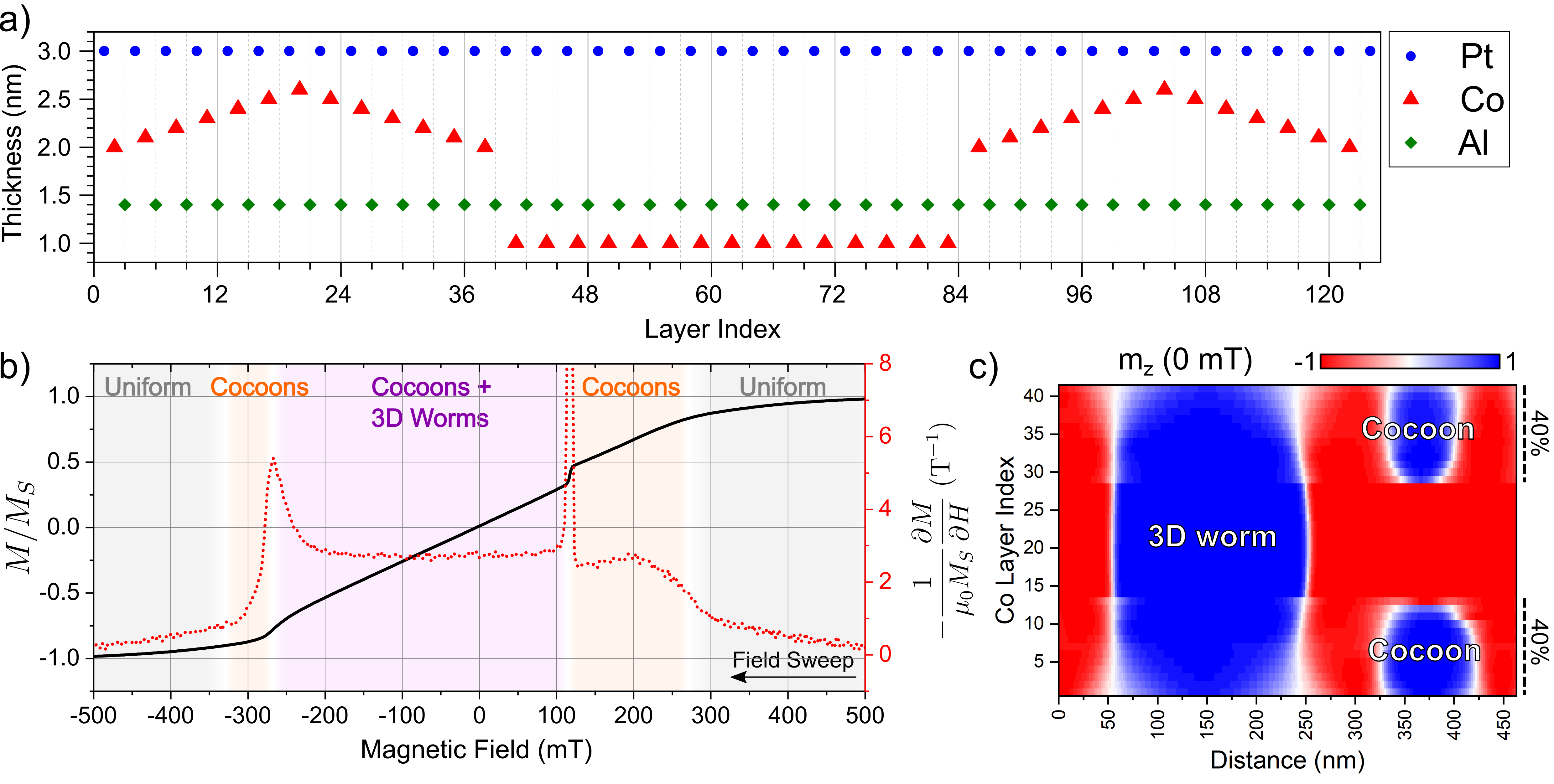}
\caption{Properties of the magnetic multilayer with variable ferromagnetic layer thickness. a) Thickness distribution of the Pt/Co/Al trilayers for each layer. b) Normalized magnetization measured via AGFM along with its derivative. The background colors identify the various expected magnetic phases expected from the derivative transitions. c) Cut of $m_z$ over the thickness showing the simultaneous presence of skyrmionic cocoons along with columnar textures at remanence. The indicated percentage values relate to the proportion of the total magnetization.}
\end{figure}

\subsection*{Temperature and field dependency of skyrmionic cocoons}	

To investigate these magnetic textures, HERALDO measurements have been performed at 300\,K and 80\,K while sweeping an OOP magnetic field from positive saturation to negative. As predicted numerically \cite{grelier2022three}, the spin textures present in the sample under focus shall significantly evolve while varying the magnetic field. Representative maps of the vertical component of the magnetization in the object hole are presented in Fig. 2a (the full field dependency is available in Suppl. Mater.). At 300\,K (top row), starting from near saturation at 330\,mT, skyrmionic cocoons have nucleated when lowering the field as shown with the image at 240\,mT. As the signal scales with $m_z$ averaged over the thickness, the existence of two different contrasts is indicative of the existence of various objects. The weaker contrast corresponds to a single cocoon, located either in the top or the bottom gradient layers, while the stronger one is linked to paired cocoons, one on top of the other. Based on the hysteresis measurements, when decreasing the field from positive saturation the middle layers should only accommodate domains for fields lower than 125mT, which thus prohibits the existence of full columnar skyrmions above that field. Those observations thus confirm the presence of different 3D textures. Then, at 100\,mT, large stripes begin to propagate into the field of view, corresponding to 3D stripes extending over the full thickness, which are still surrounded by paired cocoons. The measurement is not accurate enough to allow to differentiate between paired cocoons and 3D stripes as it only represents a 20\% variations in the signal but the results can be correlated with concomitant MFM and simulations. At remanence, a mixed state of 3D stripes and paired cocoons is observed. Upon the decrease of the magnetic field, the stripes of up magnetization get narrower and paired cocoons emerge in their stead. Finally, at -270\,mT, we can observe a few sparse skyrmionic cocoons along with a few columnar skyrmions, near the bottom left for instance. \\

\begin{figure}[htb!]
\includegraphics[width = 1\textwidth]{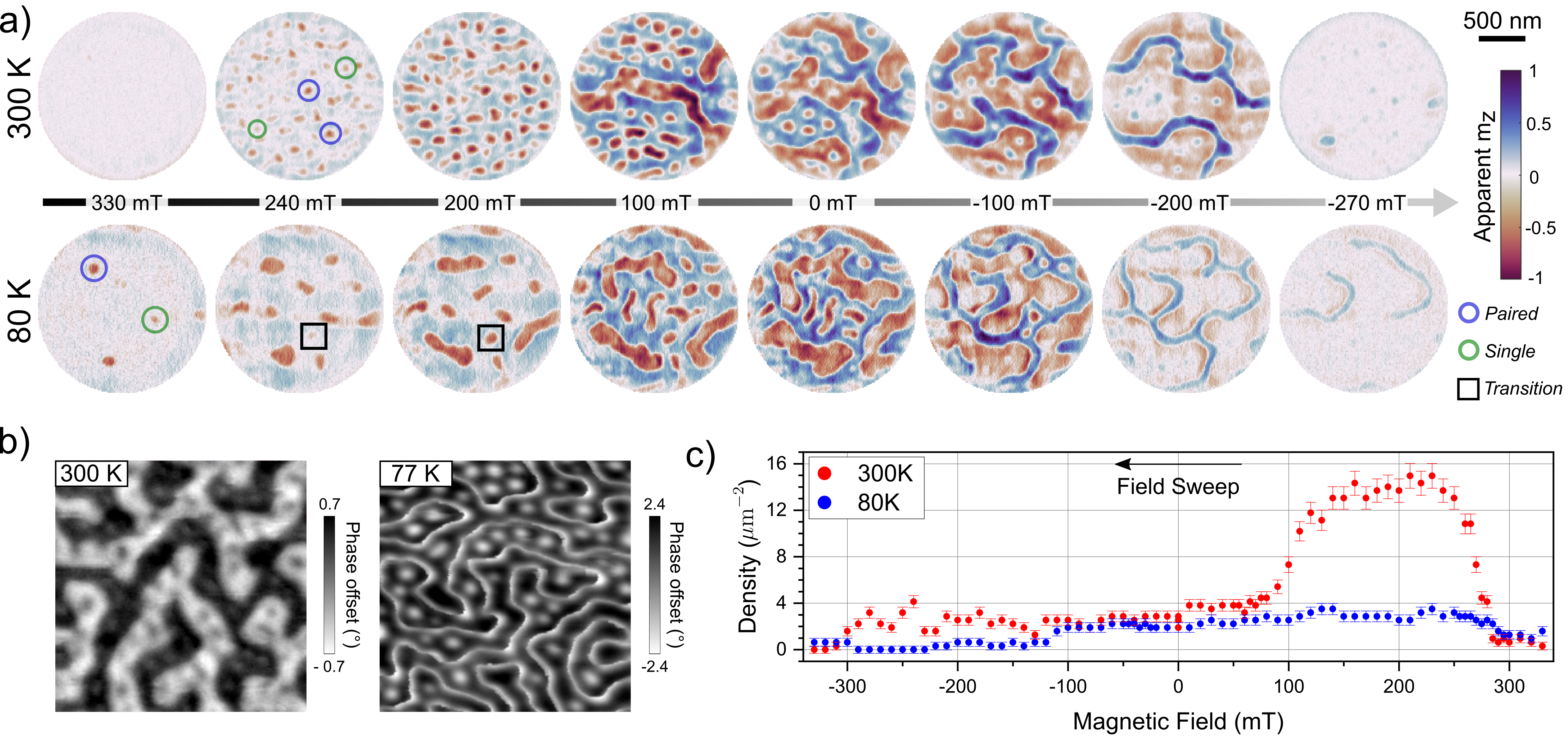}
\caption{HERALDO imaging performed with a 2 µm object hole. a) Field dependency measured at 300 (top row) and 80\,K (bottom row). The colored shapes put forward examples of paired or single cocoons cocoons as well as transitions between the two states. b) MFM images ($\rm{3\times3\ \mu m^2}$) at remanence at room temperature (left) and 77K (right). The length scale is common to a). c) Evolution of the density of isolated objects based on HERALDO measurements. }
\end{figure}

At 80\,K (bottom row in Fig. 2a), the overall evolution is similar, however a few discrepancies can be noted. First, a field of 330\,mT is not sufficient to saturate the magnetization. Notably, at positive fields, the cocoons appear to be larger and fewer in number than at room temperature. A transition between a single cocoon and a paired one is visible between the 2$\rm{^{nd}}$ and 3$\rm{^{rd}}$ frame, as indicated by the significant change in contrast in the black squares. According to magneto-transport measurements (see Suppl. Mater), the nucleation in the strong PMA layer should happen near 30\,mT. Thus at remanence, a few 3D stripes are visible as well as some elongated pairs of cocoons which correspond to the smaller objects. Other remanent states are shown in Fig. S1 (see Suppl. Mater.), showing an image at low temperature more similar to its 300\,K counterpart. Finally, decreasing the magnetic field below zero does not induce the nucleation of skyrmionic cocoons: the 3D stripes simply shrink and disappear.

To complement the holography measurements, magnetic force microscopy (MFM) measurements have been performed without an external field at room temperature and at 77K (Fig. 2b). As the latter is more sensitive to the magnetization close to the surface, it can be correlated with the holography results that relates to the average magnetization of all the magnetic layers. At 300\,K, the same characteristic pattern of stripes with smaller white dots in between is obtained. As the signal is originated from the divergence of the stray field, the weakly contrasted dots have to correspond to skyrmionic cocoons present in the top gradient of the structure and not to fully columnar skyrmions \cite{grelier2022three}. Thus, it follows that the isolated objects detected near remanence in HERALDO are paired cocoons. At low temperature, the phase signal mainly detects the domain walls which could be due to the magnetization of the tip flipping or due to an increase of the sample's anisotropy \cite{belliard1997investigation}. However, it still displays two types of structures and appear similar to its high temperature counterparts. This can be expected as no magnetic field was applied during or after the cool down so that the magnetic state was left undisturbed, \textit{i.e.} frozen in, and might be metastable. In Fig. 2c, the field dependency of the density of isolated objects is plotted for the two temperatures. At 300\,K, it is coherent with the phase diagram predicted in Fig. 1b, as the cocoons start to nucleate near 280\,mT, hence the increase in the density, and the nucleation of domains in the PMA layers near 125\,mT lowers the number of cocoons. It thus drop from nearly 15 objects per $\rm{\mu m^2}$ to oscillate around 3 objects per $\rm{\mu m^2}$ below 50\,mT. The density at high negative field depends on the way the 3D stripes disappears so variations are expected depending on the magnetic history. At 80\,K, the density does not exceed 4 objects per $\rm{\mu m^2}$ over the whole field range. The observed differences can attributed to the evolution of the magnetic parameters with the temperature. The anisotropy was determined through transport measurements (see Suppl. Mater.) while the saturation magnetization was retrieved with SQUID (see Table 1).

\begin{table}[htb!]
\begin{center}
\begin{tabular}{c c c} 
 \hline
 Temperature (K) & $K_{\rm{u}}\ {\rm{(MJ/m^{-3}})}$  & $M_{\rm{S}}\ \rm{(MA/m)}$ \\
 \hline\hline
 80 & 0.93 $\pm$ 0.03 & 1.26 $\pm$ 0.02\\ 
 300 & 0.75 $\pm$ 0.03 & 1.22 $\pm$ 0.02\\
 \hline
\end{tabular}
\caption{Magnetic parameters measured at different temperatures.}
\end{center}
\end{table}

As the skyrmionic cocoons can be stabilized with this increased anisotropy and magnetization saturation, it confirms that they can be observed over a large range of magnetic parameters and thus inside many multilayers architectures.

Thanks to the large field of view (2 µm), the object hole accommodates many different features which have been fitted, when possible, with a two-dimensional rotated Gaussian model for every field (see Methods). An example with the coexistence of isolated and paired cocoons is studied in Fig. 3a, at 300\,K and 240\,mT, as shown in the inset. The average full width at half maximum (FWHM) and the amplitude of the Gaussian for each objects are displayed in the histograms and the right panel shows a typical fit that was performed on a cocoon. The average sizes are scattered on a large range, starting as small as 50\,nm up to 130\,nm diameter. Similarly, the amplitudes spans different values, the lowest corresponds to single cocoons whereas the highest are more likely to be associated with paired cocoons. Then for each magnetic field, the quartiles of the average FWHM were extracted and the results are presented in Fig. 3b at 300 and 80\,K, where the 1$\rm{^{st}}$ and 3$\rm{^{rd}}$ quartiles serve as error bars to frame the median value while the color encodes the density. The characteristic diameter at low temperature displays a stronger dispersion, reaching sizes above 200\,nm whereas at room temperature it mostly stays below 130\,nm. In both cases, we can notice the almost linear increase of the size when decreasing the field from positive saturation.\\

\begin{figure}[htb!]
\includegraphics[width = 1\textwidth]{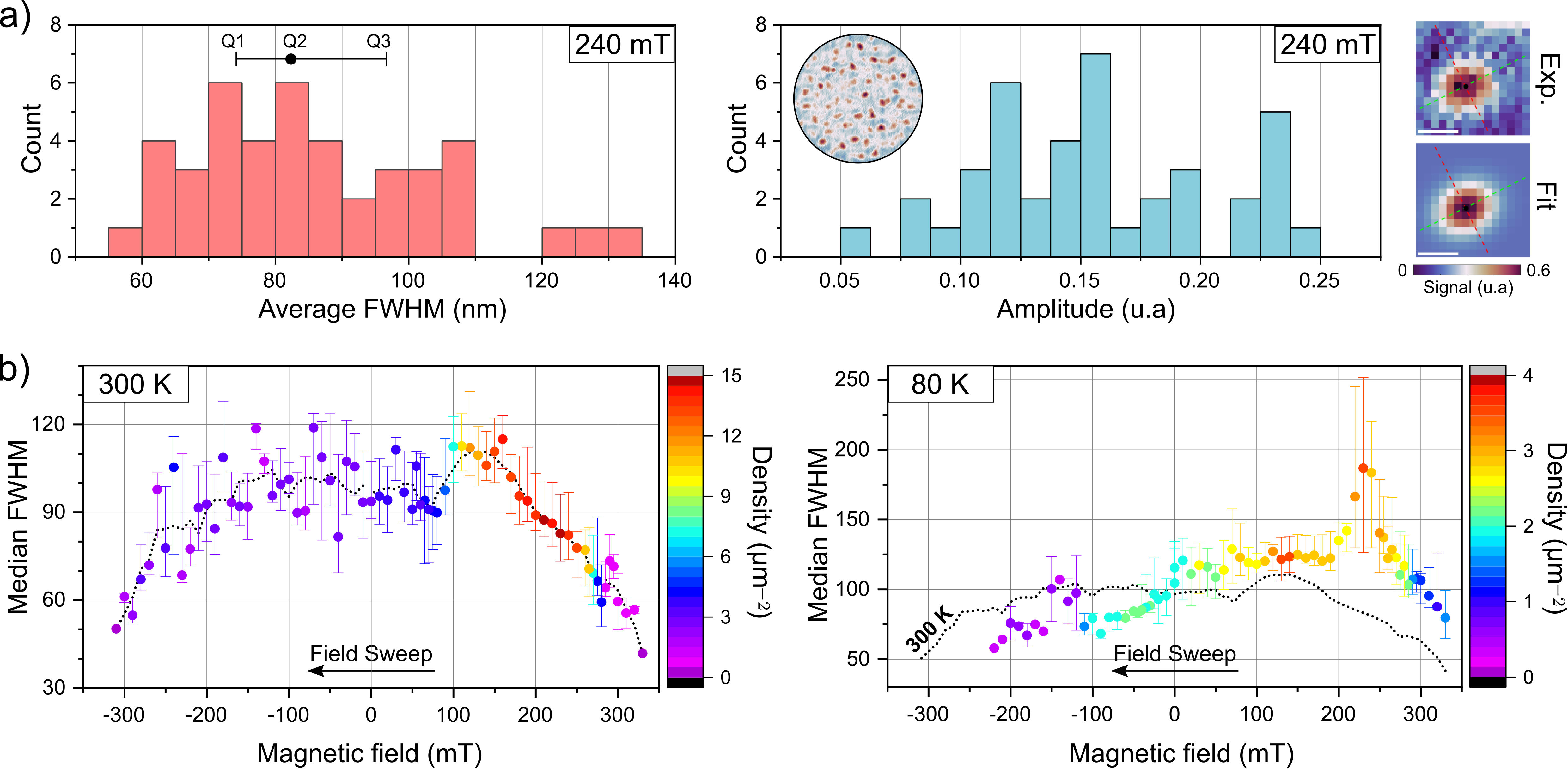}
\caption{Size study of the isolated objects observed with HERALDO based on two-dimensional Gaussian fits. a) Size and amplitude distribution at 240\,mT. The images at the right display an example of experimental data along with the corresponding fit. The two dotted lines respectively correspond to each of the FWHMs which are then averaged for each object to yield the size histogram. The white scale in both images is 60\,nm long and pixels are $\rm{15\times15\ nm^2}$. The inset image shows the magnetic state in the 2 µm object hole with a reduced colorscale. b) Field evolution of the median FWHM (Q2) along with the color coded density at 300 and 80 K. The error bars correspond to the first (Q1) and third quartile (Q3) of the average FWHM as illustrated in a).}
\end{figure}

\subsection*{Magnetic phase transitions and interactions}

As an alternative measurement, the second sample with a slightly different Pt distribution was probed using FTH through an 800\,nm object hole. The recorded holograms have been reconstructed making use of phase retrieval algorithms to improve the contrast and resolution (see Methods for details). For a better visual display, a bilinear interpolation has been performed on the reconstructions, increasing the number of pixels by a factor 1.25. In Fig. 4a, FTH reconstructed real-space images of the object hole are displayed at different OOP fields starting from a demagnetized state, prepared with a field tilted at 45$^{\circ}$ with respect to the sample normal. At first, only two dark red 3D stripes are visible. As the field increases, they get narrower and some cocoons, with a weaker contrast, start to appear at 260\,mT. At 320\,mT, the stripes have transformed into columnar skyrmions, still accompanied by single cocoons. Furthermore, the images at 342 and 348\,mT put forward the transition between a columnar skyrmion and two vertically aligned cocoons (as highlighted with the pink circles) which have a stronger contrast than the single cocoon also visible in those frames. While it was not possible to differentiate the two states at remanence, the difference at high field is significant enough to classify them. This is also supported by the numerical prediction of this type of transition \cite{grelier2022three}.

In Fig. 4b, we start from near saturation with a few skyrmionic cocoons and lower the magnetic field which allows the nucleation of more skyrmionic cocoons, as shown with the image measured at 327\,mT.  The density continues to rise, as expected from the previous study, until the cocoons begin to interact. Looking closer at the transitions between 306 and 280\,mT, the ‘merging’ of two pairs of skyrmionic cocoons can be observed, as evidenced by the colored ellipses. Indeed, the cocoons appear to be attracted to one another, implying that one resides in the bottom gradient and the other one in the top one. This transition, which places them one on top of the other, allows to reduce the dipolar interaction, thus reaching a minimum of energy. The absence of more transitions of this kind at higher fields is most likely due to pinning of the cocoons caused by inhomogeneities in the magnetic interactions associated to the material's grains formed during growth (see Suppl. Mater. and Refs. \cite{legrand2017room,kim2017current,marioni2018halbach}). Only when the field is lowered, the cocoons expand and increase their dipolar energy until it is sufficiently strong to overcome the pinning, thus allowing their alignment. As the field decreases further (for instance at 208 mT), additional aligned cocoons appear and become more elongated. At remanence, we observe again a 3D worm surrounded by paired cocoons, as supported by MFM and micromagnetic simulations.

In Fig. 4c, profiles are shown at different fields to illustrate the transition between the single cocoon state to the double one, evidencing the associated change in contrast. Interestingly, at first, only one of the cocoons is moving towards the other (see black arrow on the left graph) which suggest that the latter is pinned. However, once merged, the two aligned cocoons have slightly moved away from their previous position, a motion likely due to interactions with the other magnetic objects. This peculiar attractive interaction as well as the change in contrast confirms the 3D nature of the skyrmionic cocoons.

\begin{figure}[htb!]
\includegraphics[width = 1\textwidth]{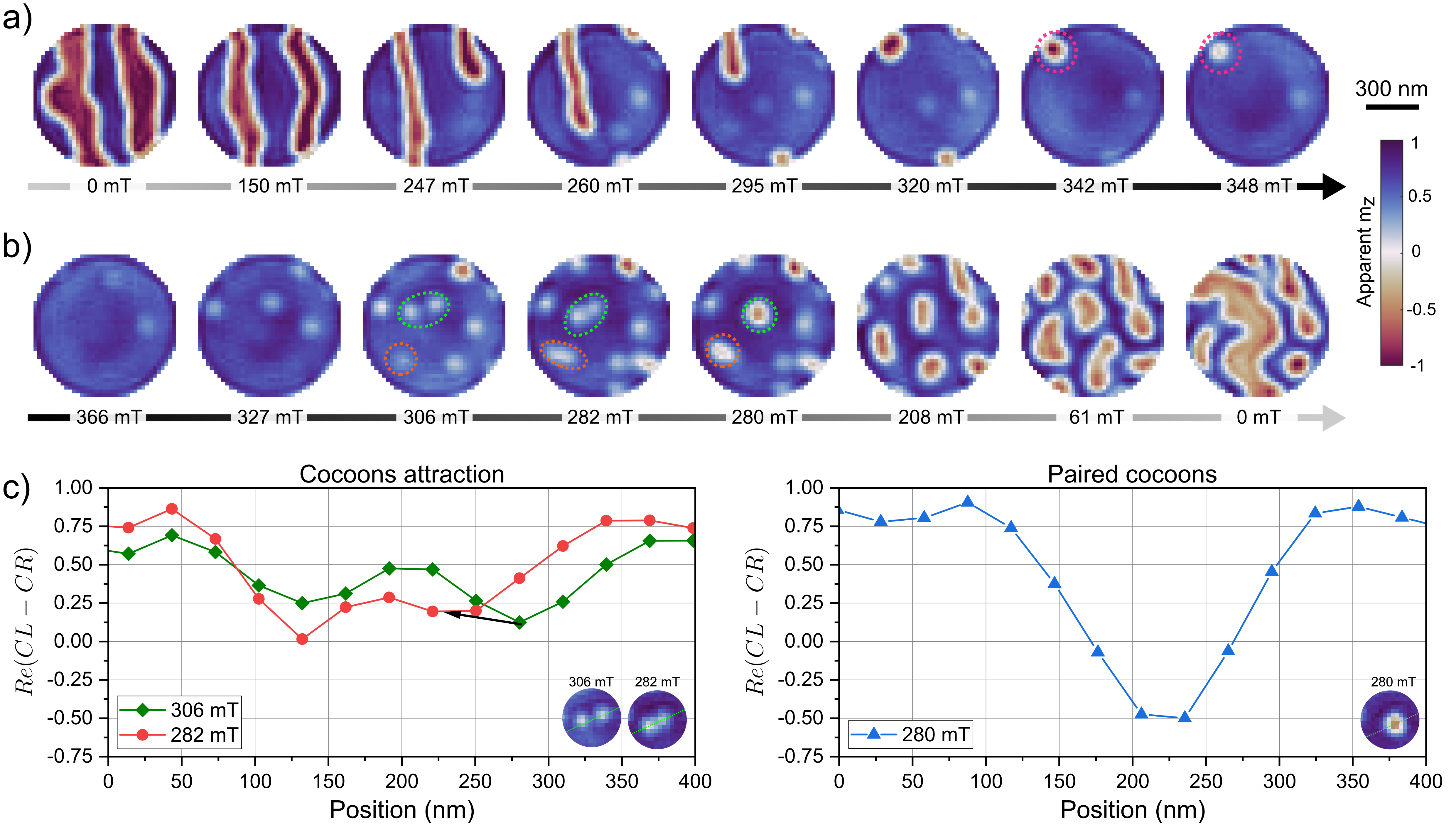}
\caption{FTH imaging performed with a 800\,nm object hole. Dependency with a) increasing and b) decreasing magnetic field. The colored ellipses follow the same textures over different fields to highlight the associated transitions. c) Study of the transition between two isolated cocoons. The black arrow indicates the relative motion of the cocoons between the two successive fields. The position of the linecut has been changed between the two graphs but the inset shows the same sample area (a circle of 200\,nm radius).}
\end{figure}

\section*{Conclusion}

The X-ray holography experiments demonstrate the presence of various 3D magnetic textures in aperiodic multilayers. Additionally, it allows an in-depth study of the field-dependent behavior of skyrmionic cocoons which evidenced complex interactions between them, mostly governed by the dipolar interaction. Those effects are strong enough to move the skyrmionic cocoons, by a typical distance of the order of 50\,nm, so that they can coalesce by pair. It also put forward the various magnetic phases that undergoes the magnetization under an external field as well as the associated phase transitions which were captured by those measurements. The temperature study highlight a large range of stability for the skyrmionic cocoons even though their size and density may vary. The holography precision limit regarding the contrast does not systematically allow us to distinguish between fully columnar textures and two vertically aligned cocoons, as it only represents a difference of 20\% of the total magnetization. However, the input from MFM and micromagnetic simulations allows to confidently determine their actual nature. This study thus establishes that 3D skyrmionic cocoons can reside in different vertical position of the multilayer, i.e. in the bottom and the top gradient, which could serve as a first basis towards the development of a 3D memory device.
\vspace{1\baselineskip}

\paragraph*{\textbf{Acknowledgments}}
Financial supports from FLAG-ERA SographMEM (ANR-15-GRFL-0005), from ANR under the grant ANR-17-CE24-0025 (TOPSKY), ANR-20-CE42-0012 (MEDYNA) and ANR-22 DFG (TOPO3D) and as part of the ‘Investissements d’Avenir’ program SPiCY (ANR-10-LABX-0035). R.B. and F.B. acknowledge funding from the Helmholtz Young Investigator Group Program. We would like to thank M. Bonnet and M. Vallet for access and operation of the FIB to prepare the HERALDO sample.

\section*{Methods}

\paragraph*{Sample preparation and characterization}
\indent The samples have been grown on thermally oxidized silicon substrates using magnetron sputtering at room temperature. A seed layer of 5\,nm Ta is typically used and 3\,nm Pt capping is deposited to protect from oxidation. In details, the HERALDO sample corresponds to \seqsplit{Ta5|Pt3|(Co[2.0:0.1:2.5]|Al1.4|Pt3)(Co[2.6:0.1:2.0]|Al1.4|Pt3)|(Co1.0|Al1.4|Pt3)}$\times$\seqsplit{15|(Co[2.0:0.1:2.5]|Al1.4|Pt3)(Co[2.6:0.1:2.0]|Al1.4|Pt3)} where the notation $\left[X_1:S:X_2\right]$ corresponds to the thickness sequence between $X_1$ and $X_2$ with $S$ being the thickness step. The FTH sample is similar with only a difference in the Pt thickness in the last layers: \seqsplit{Ta5|Pt3|(Co[2.0:0.1:2.5]|Al1.4|Pt3)(Co[2.6:0.1:2.0]|Al1.4|Pt3)|(Co1.0|Al1.4|Pt3)}$\times$\seqsplit{15|(Co[2.0:0.1:2.5]|Al1.4|Pt3)(Co[2.6:0.1:2.0]|Al1.4|Pt2) (Co2.0|Al1.4|Pt3)}.

The multilayers have been grown on the top side of a Si$\rm{_3}$N$\rm{_4}$ membrane. For the first (resp. second) sample, the back side was covered with a 1.1 µm-thick gold mask layer  (Al 2.5\,nm/Au 25\,nm)$\times$40 (resp. (Cr 5\,nm/Au 55\,nm)$\times$20) that was milled away with focused ion beam to form a 2 µm (resp. 800\,nm) object hole. For FTH, reference holes of different sizes, typically 30-40\,nm, were punched through the whole structure at a distance of ~1.8 µm from the object hole center. For HERALDO, the slit used for the reconstructions was 5 $\rm{\mu m}$ long and 30\,nm wide.

The magnetic hysteresis have been measured using AGFM, and SQUID to determine the saturation magnetization at 300 and 80\,K.

\paragraph*{Magnetic force microscopy and demagnetization}
\indent Before imaging, the samples have been demagnetized using an electromagnet which applied an oscillating magnetic field of exponentially decreasing amplitude. At room remperature, the MFM images have been acquired using cantilevers from TeamNanotech (TN) with nominal stiffness of 3\,N/m, with its tip capped by 7\,nm of magnetic (proprietary) material and 10\,nm of Pt. We used a tapping mode with a lift height of 10\,nm, at 75\% of the drive amplitude with double-passing. The measurements at low temperature have been performed in a liquid nitrogen bath with 15 mbar of Helium as the exchange gas, using a TeamNanotech tip with a 0\,nm lift height and at 50\% of the drive amplitude.

\paragraph*{FTH reconstruction - Phase Retrieval}
	
\indent The holograms were recorded for left (CL) and right (CR) polarization of the X-ray light in identical conditions at the Co L$\rm{_3}$ and then pre-processed for the iterative phase retrieval algorithm. The polarization level of the light is superior to 99\%. Left- and right-polarization hologram were first normalized with respect to each other, centered and any constant offset present was removed. The Fourier transform (FT) of this modified holograms yields a crude real space image of the cross correlation of the reference holes with the object holes.

The phase retrieval process consists on digitally propagating the reconstructed image back and forth from the reciprocal to the real space, while imposing two constraints. In real space, the support constraint forces the reconstructed image to be located only in the regions were light is actually transmitted (the support is manually constructed using the FTH reconstructions as a reference), and in reciprocal space the amplitude of the propagated hologram is forced to match the experimentally recorded one, until the algorithm converges to a final reconstruction respecting both constraints. Pixels of the recorded hologram carrying unreliable values (e.g. the pixels shadowed by the beamstop used to protect the camera from the direct beam) were not subjected to the amplitude constraints and their value was left free to change.

Once the data was pre-processed, we began by reconstructing the CL image with a combination of the oversampling smoothness (OSS) algorithm \cite{rodriguez2013oversampling} (250 iterations) followed by the solvent flipping (SF) algorithm  \cite{marchesini2007invited} (200 iterations), which typically yielded a relative error with respect to the experimental hologram of 0.1\%. The convergence parameter $\beta$ started from 0.9 and decreased progressively with an \textit{arctan} trend down 259 to 0.5 over the OSS iterations. The filter parameter $\alpha$ decreased linearly from the image size down to 0. Then, the solution for CL polarization was used as a starting guess for the CR polarization image which thus only went through the SF algorithm (200 iterations).
For each polarization, an average of the 20 reconstructions with the lowest errors was extracted. From that, magnetic contrast is extracted by computing the difference of the two polarized images.

The dark circular background visible in some of the images is an artifact arising from erroneous reconstruction of the missing signal at low frequencies, corresponding to pixels covered by the beamstop for which not only the phase, but also the signal amplitude had to be reconstructed.

\paragraph*{HERALDO}

\indent Holograms were recorded at the Co L$\rm{_3}$ edge for both helicities of the X-ray light. The beamstop is smoothed using the function introduced in \cite{streit2009magnetic}. After normalization and centering, a differential filter was applied to correct one slit while the other one was set to 0 to cancel the corresponding periodic stripes in the reconstruction. After a Fourier Transform of the difference that yields the magnetic information, the phase of the real space image is optimized to maximize the contrast, giving the final image. A gaussian filter with 0.5 standard deviation was applied to display the final images (not for the analysis).

The image analysis was performed under Matlab, using binarization and edge detection to automatically identify isolated objects. Shapes having an aspect ratio bigger than 2.5 and a circularity $C=4\times\pi\times Area/Perimeter^2$ smaller than 0.2 were discarded during the process to eliminate artefacts and elongated stripes. The remaining objects allow to compute the density. Finally, they were fitting using a rotated 2D Gaussian model:

\begin{equation}
z(x,y) = z_0 + A\exp{\left[-\left(\dfrac{\left(x-x_0\right)\times \cos{\theta}-\left(y-y_0\right)\times \sin{\theta}}{\sigma_1}\right)^2-\left(\dfrac{\left(x-x_0\right)\times \sin{\theta}+\left(y-y_0\right)\times \cos{\theta}}{\sigma_2}\right)^2\right]}
\end{equation}

With $x_0$, $y_0$ and $z_0$ the offsets, $A$ the amplitude and $\sigma_{i}$ the standard deviations linked to the FWHM with:

\begin{equation}
\rm{FWHM}_i = 2\sqrt{\rm{log}(2)}\times \sigma_i
\end{equation}

\paragraph*{Micromagnetic simulations}

The micromagnetic simulations were performed using the Mumax3 solver \cite{vansteenkiste2014design}. At room temperature, the parameters were fixed as follow: the exchange constant $A = 18\ \rm{pJ.m^{-1}}$, the saturation magnetization $M_S = 1.2\ \rm{MA.m^{-1}}$, the uniaxial interfacial anisotropy constant $K_{u,s} = 1.62\ \rm{mJ.m^{-2}}$ and the interfacial DMI constant $D_s = 2.34\ \rm{pJ.m^{-1}}$. Given the complex structure and the variable thickness, each trilayers was divided in three layers, one magnetic and two empty and the magnetic parameters were diluted depending on the experimental thickness \cite{lemesh2017accurate}. The total simulation space was made of $512\times 512\times 121$ cells, each of size $2 \times 2 \times 1.8\ \rm{nm^3}$.


\end{document}